\documentclass[aps,prb,twocolumn,superscriptaddress]{revtex4-1}
\usepackage{graphicx} % For importing images
\usepackage{float} % extra commands for placing figures
\usepackage{comment} % allows to comment out text
\usepackage{amsmath} % makes maths look nice

\newcommand{\csfese}{$Cs_{0.8}Fe_{1.6}Se_2$}	%Cs0.8Fe1.6Se2  \csfese
\newcommand{\rtF}{$\sqrt{5}\times\sqrt{5}$}      % rt5xrt5    	\rtF
\newcommand{\rtT}{$\sqrt{2}\times\sqrt{2}$}      % rt2xrt2    	\rtT

\begin{document}

\title{Two-dimensional Cs-vacancy superstructure in iron-based superconductor $Cs_{0.8}Fe_{1.6}Se_2$}

\author{D.G. Porter}
\email[]{dan.porter@diamond.ac.uk}
\affiliation{Department of Physics, Royal Holloway, University of London, Egham, TW20 0EX, UK}
\altaffiliation{Present address: Diamond Light Source Ltd, Harwell Science and Innovation Campus, Didcot, Oxfordshire, OX11 0DE, UK}
\author{E. Cemal}
\affiliation{Department of Physics, Royal Holloway, University of London, Egham, TW20 0EX, UK}
\affiliation{Institut Laue-Langevin, BP156, F-38042 Grenoble Cedex 9, France}
\author{D.J. Voneshen}
\affiliation{Department of Physics, Royal Holloway, University of London, Egham, TW20 0EX, UK}
\author{K. Refson}
\affiliation{Department of Physics, Royal Holloway, University of London, Egham, TW20 0EX, UK}
\affiliation{ISIS, Science and Technology Facilities Council, Rutherford Appleton Laboratory, Didcot OX11 0QX, UK}
\author{M.J. Gutmann}
\affiliation{ISIS, Science and Technology Facilities Council, Rutherford Appleton Laboratory, Didcot OX11 0QX, UK}
\author{A. Bombardi}
\affiliation{Diamond Light Source Ltd, Harwell Science and Innovation Campus, Didcot, Oxfordshire, OX11 0DE, UK}
\author{A.T. Boothroyd}
\affiliation{Clarendon Laboratory, Parks Road, Oxford OX1 3PU, UK}
\author{A. Krzton-Maziopa}
\affiliation{Faculty of Chemistry, Warsaw University of Technology, PL-00664 Warsaw, Poland}
\author{E. Pomjakushina}
\affiliation{Laboratory for Developments and Methods, Paul Scherrer Institute, CH-5232 Villigen PSI, Switzerland}
\author{K. Conder}
\affiliation{Laboratory for Developments and Methods, Paul Scherrer Institute, CH-5232 Villigen PSI, Switzerland}
\author{J.P. Goff}
\affiliation{Department of Physics, Royal Holloway, University of London, Egham, TW20 0EX, UK}

\date{\today}

\begin{abstract}
Single crystal neutron diffraction is combined with synchrotron x-ray scattering to identify the different superlattice phases present in $Cs_{0.8}Fe_{1.6}Se_2$. A combination of single crystal refinements and first principles modelling are used to provide structural solutions for the $\sqrt{5}\times\sqrt{5}$ and $\sqrt{2}\times\sqrt{2}$ superlattice phases. The $\sqrt{5}\times\sqrt{5}$ superlattice structure is predominantly composed of ordered Fe vacancies and Fe distortions, whereas the $\sqrt{2}\times\sqrt{2}$ superlattice is composed of ordered Cs vacancies. The Cs vacancies only order within the plane, causing Bragg rods in reciprocal space. By mapping x-ray diffraction measurements with narrow spatial resolution over the surface of the sample, the structural domain pattern was determined, consistent with the notion of a majority antiferromagnetic $\sqrt{5}\times\sqrt{5}$ phase and a superconducting $\sqrt{2}\times\sqrt{2}$ phase.
\end{abstract}


\maketitle

% Section 1
\section{Introduction}
The alkaline iron selenides $A_xFe_{2-y}Se_2$ ($A =$ K, Rb, Cs) have attracted much interest recently due to the observation of superconductivity with $T_c \approx 30K$\cite{Guo2010,Wang2011c,KrztonMaziopa2011,Fang2011} in conjunction with antiferromagnetism with an unusually high ordering temperature $T_N$ of up to 559K and large ordered moment of about 3.3$\mu_B$ per Fe\cite{Bao2011}. This observation suggests the possibility of coexistence between these two orthogonal phenomena, superconductivity and magnetic ordering. Other iron pnictide superconductors have phase diagrams that indicate coexistence of magnetism and superconductivity in certain regions, however the highest $T_c$'s and bulk superconductivity are only found when the magnetic state has been suppressed\cite{Johnston2010,Stewart2011,Lumsden2010}. The basic principles of superconductivity indicate that magnetic fields cannot permeate a superconducting region.

At elevated temperatures these compounds exhibit the tetragonal space group $I4/mmm$. Fe vacancies order below $\approx$ 600K into a \rtF{} structure with $I4/m$ symmetry\cite{Zavalij2011}. The Fe lattice in this structure magnetically orders into a block antiferromagnet, with neighbouring groups of four Fe ions having opposite spins\cite{Bao2011}. Other phases have been observed in the system by diffraction and scanning electron microscopy, indicating the presence of a \rtT{} phase\cite{Bosak2012,Ricci2011a} and even a $2\times4$ rhombus phase\cite{Zhao2012}. This has led to the idea that the superconductivity in this system may be in a different phase to the principal magnetic phase. 

% Additional paragraph of introduction for Ref 2, comment 1. 20 March 2015
The observation of multiple phases has driven a number of reports on phase-separation in this system, using various techniques to probe either the surface or the bulk material. Transmission electron microscopy (TEM) measurements observed parallel lamellae of ordered and disordered Fe-vacancy regions\cite{Wang2011b} and synchrotron x-ray diffraction measurements determined the nanoscale phase-separation of a magnetic, Fe-vacancy superlattice and a nonmagnetic phase with an in-plane compressed lattice\cite{Ricci2011}. Scanning Tunneling Microscope (STM) measurements showed that the superconducting state would not survive in regions with Fe vacancies\cite{Li2011} and finally bulk measurements with NRM concluded that the superconducting and antiferromagnetic properties are part of two different phases\cite{Texier2012}.

In this paper we will describe the results of single crystal diffraction measurements on \csfese{} leading to the determination of the \rtT{} superstructure and show that this phase is spatially separated from the magnetic \rtF{} structure by scanning the sample surface with a focused x-ray beam.
 
% Section 2
\section{Methods}
% Changed Reference YuPomjakushin2012 to Pomjakushin2011 for Ref 1, comment 1. 20 March 15
% Added line "In summary, ... T_N~500K" for Ref 1, comment 1. 20 March 15
% Specified "second sample from the same crystal growth" for Ref 2, comment 2.
Single crystals of $Cs_xFe_{2-y}Se_2$ were grown by the Bridgman process as described in Ref. \cite{KrztonMaziopa2011}. The nominal composition of the crystals used in this study is \csfese{}, and their superconducting and magnetic properties have been reported previously\cite{KrztonMaziopa2011,Pomjakushin2011,Taylor2012,Speller2012}. In summary, the inclusion of Cs as the intercalation ion leads to a large increase in the $c$-lattice parameter compared with $K_xFe_{2-y}Se_2$, despite this however, the reported superconducting transition is roughly the same at $T_c \approx 30K$ and the antiferromagnetic ordering transition is still above room temperature at $T_N \approx 500K$. The samples were coated in Cytop varnish to protect them from reaction with the atmosphere. A single grain sample was measured using a white beam of neutrons on the SXD instrument installed at the ISIS spallation neutron source (Oxfordshire, UK) for single crystal diffraction analysis\cite{Keen2006}. To determine the spatial separation of the domains, a second sample from the same crystal growth was cleaved under a He atmosphere and measured with a focused beam of synchrotron x-rays on beamline I16 at Diamond light source (Oxfordshire, UK)\cite{Collins2010}. 

Density Functional Theory (DFT) calculations were performed using the plane-wave pseudopotential approach as implemented in the CASTEP code\cite{Clark2005}. The initial calculations used the block antiferromagnetic structure previously shown to be the lowest energy state for systems with no Fe vacancies\cite{Li2012DFT}. A system with two layer periodicity was geometry optimised such that the residual forces were $|F| <= 0.036 eV/A$. This optimised structure was then used as the base for all further calculations. Changes to the Cs layer position were not found to increase the maximum residual force in any layer and consequently no further optimisations were performed. The spin polarised Perdew-Burke-Ernzerhof generalised gradient approximation was used to model the exchange and correlation\cite{Perdew1996}. The default CASTEP ultrasoft pseudopotentials (Version 7) were used with a plane wave cutoff of 550 eV. The electronic Brillouin zone was sampled by a Monkhorst-Pack grid\cite{Pack1977} with a density of at least 0.02 \AA. Calculations were performed with a Gaussian smearing of 0.05 eV, giving a total error in the calculation of energy of about 0.2 meV.

% Section 3
\section{Single Crystal Neutron Diffraction}
\subsection{Parent Structures}
\begin{figure}[ht]
	\centering
	\includegraphics[width=0.5\textwidth]{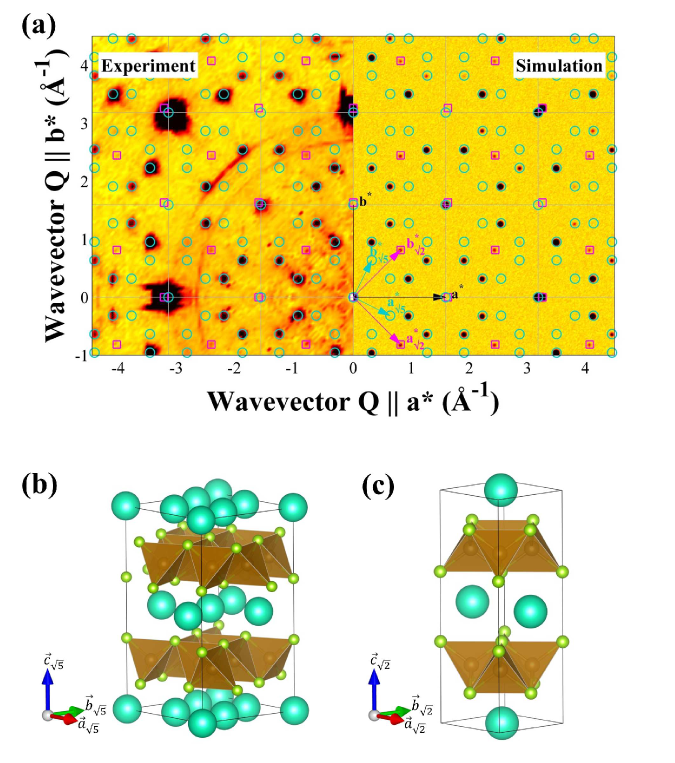}
	\caption{(Colour online) Single crystal diffraction pattern of \csfese{} measured on SXD. A cut through the $(h,k,0)$ plane is illustrated on the left in (a) with superlattice peaks visible from both phases present in the sample: the \rtF{} phase illustrated in (b) and the \rtT{} phase in (c). Light blue circles show the \rtF{} lattice, light red squares the \rtT{} lattice. A simulation of the diffraction data on the right hand side, using the models in (b) and (c), exhibiting good agreement with the measured data.} \label{fig:SXDhk0}
\end{figure}

A single crystal of \csfese{} was mounted in a low background sample environment on SXD and measured at the base temperature, 30K. Several orientations were measured to build complete coverage of reciprocal space and the orientation was chosen so that the principal axes were centred in the rear detectors with highest resolution in wavevector transfer. The resulting diffraction pattern, illustrated in Fig. \ref{fig:SXDhk0}, shows a large number of reflections. The larger peaks are from two overlapping tetragonal average structures and the smaller peaks are satellites produced by commensurate superstructures. Concentrating initially on the principal reflections of the average structures, the two overlapping phases have slightly different lattice parameters. The lattice parameters were determined by fitting the peak profiles along the principal directions when these directions were aligned at the centre of a back scattering detector. Phase 1 has the greatest intensity and lattice parameters $a_1=3.9472(2)$\AA{} and $c_1=15.2401(1)$\AA{}. The weaker second phase is expanded along the $c$ direction and contracted in $a$, having $a_2=3.8554(1)$\AA{} and $c_2=15.4931(5)$\AA{}. The variation of the peak intensity indicates the sample fraction of phase 2 to be $42(3)\%$.

% Figure 1
% Added "Experiment" & "Simulation" for Ref 1, comment 2. 20 March 15

% Figure 2
% Added "Experiment" & "Simulation" for Ref 1, comment 2. 20 March 15
\begin{figure}[ht]
	\centering
	\includegraphics[width=0.47\textwidth]{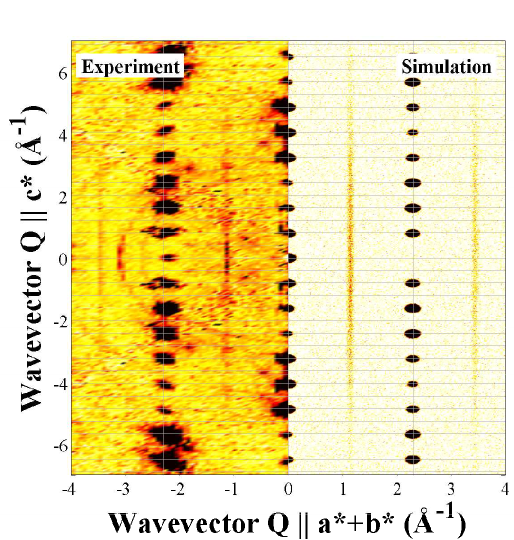}
	\caption{(Colour online) Bragg rods observed in the $(h,h,l)$ plane of the neutron diffraction data. The left-hand side gives the neutron diffraction and the right-hand side compares the simulated data from a model composed of 100 randomly varying iterations of the Cs-vacancy structure, approximating disorder along $c$.} \label{fig:SXDhhl}
\end{figure}

% Figure 3
% Added inset figures of reflections with Q<4A and updated caption for Ref 2, comment 4. 20 March 15
\begin{figure}[ht]
	\centering
	\includegraphics[width=0.47\textwidth]{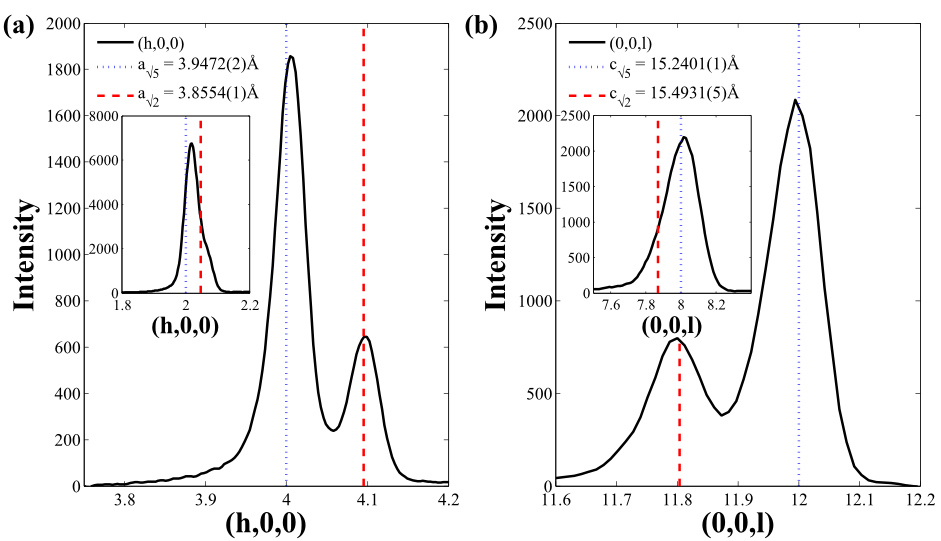}
	\caption{(Colour online) Separation of Bragg reflections along principal directions in single crystal neutron diffraction data. In each case the peak is split into two components; a larger component with larger $a$ and smaller $c$ and a weaker component. Positions of each set of lattice parameters for phase 1 and 2 are shown by blue and red dashed lines respectively. Inset within each plot is an example of a reflection in the same direction with $|Q|<4$\AA$^{-1}$, where the peaks are strongly overlapping.} \label{fig:SXDh00}
\end{figure}

These large Bragg peaks were integrated using 3D profile fitting methods and refined using the program Jana2006\cite{Petricek2014}. The two phases were indexed, integrated and refined independently after absorption corrections were applied. Refining the average structure is unable to give information on the ordering of vacancies but can be used to determine the average occupancies and thermal parameters of different sites. Initially the refinements from both phases gave similar results and were consistent with previous literature on this phase\cite{Zavalij2011}, with Cs and Fe concentrations of approximately 80\% of the fully occupied values. It was realised, however, that a number of reflections overlap between the two phases which would have an averaging effect on the occupancies. At low wavevector transfers, where reflections are closer together in reciprocal space, overlapping reflections are more likely due to the low detector resolution on SXD. All peaks with low wavevector transfer were omitted and a cut off of $|Q|<4$\AA$^{-1}$ was used. The limit of $|Q|<4$\AA$^{-1}$ was chosen as it was possible to distinguish separate profiles for overlapping reflections above this limit, as can be seen in Fig. \ref{fig:SXDh00}. Refinements from the resulting reflections are presented in Table \ref{tab:RTtwo}. 

% Table 1
% Caption updated to explain Refs and give max |Q| for Ref 2, comment 3. 20 March 15
\begin{table}[ht]
\centering
\begin{tabular}{r|cccccc} 
\multicolumn{6}{c}{\textit{(a) All reflections}} \\
              & Atom & Site & Occupancy & $U_{11/22}$ & $U_{33}$ \\ \hline
Phase 1       & Cs   & $2b$ & 0.82(6)   & 0.036(3) & 0.007(3) \\
Refs: 681     & Fe   & $4d$ & 0.81(1)   & 0.0100(6) & 0.0033(8) \\
$R_w=13.89\%$ & Se   & $4e$ & 1         & 0.0116(6) & 0.0093(8) \\ \hline
Phase 2       & Cs   & $2b$ & 0.83(6)   & 0.039(6) & 0.021(7) \\
Refs: 480     & Fe   & $4d$ & 0.80(2)   & 0.010(1) & 0.007(2) \\
$R_w=23.89\%$ & Se   & $4e$ & 1         & 0.013(1) & 0.017(2) \\ 
\multicolumn{6}{c}{\textit{(b) Only reflections with $|Q|>4$\AA$^{-1}$}} \\
              & Atom & Site & Occupancy & $U_{11/22}$ & $U_{33}$ \\ \hline
Phase 1       & Cs   & $2b$ & 0.97(5)   & 0.042(4) & 0.014(3) \\
Refs: 552     & Fe   & $4d$ & 0.81(1)   & 0.010(1) & 0.004(1) \\
$R_w=13.07\%$ & Se   & $4e$ & 1         & 0.013(1) & 0.010(1) \\ \hline
Phase 2       & Cs   & $2b$ & 0.63(9)   & 0.023(7) & 0.012(8) \\
Refs: 381     & Fe   & $4d$ & 0.96(4)   & 0.012(2) & 0.010(2) \\
$R_w=24.47\%$ & Se   & $4e$ & 1         & 0.011(2) & 0.013(2) \\
\end{tabular}
\caption{Occupancies and anisotropic thermal parameters of the two phases of \csfese{} by refinements of the principal Bragg reflections. Both phases are refined in the tetragonal space group $I4/mmm$. Phase 1 has lattice parameters $a_1=3.9472(2)$\AA{}, $c_1=15.2401(1)$\AA{} and phase 2 $a_2=3.8554(1)$\AA{} and $c_2=15.4931(5)$\AA{}. The atomic sites are $2b = (\frac{1}{2},\frac{1}{2},0)$, $4d = (\frac{1}{2},0,\frac{1}{4})$, and $4e = (0,0,0.1540(3))$. Reflections were collected up to a maximum $|Q|=16$\AA$^{-1}$, the number of reflections observed in each dataset is given by ``Refs:''. Table (a) gives the refined values when all measured reflections are used and table (b) shows the improved values given when refined using high resolution reflections.}
\label{tab:RTtwo}
\end{table}

When low resolution reflections are omitted, the occupancies of the two phases become different. The first phase refines to a concentration of $Cs_{0.97(5)}Fe_{1.60(4)}Se_2$, consistent with a filled Cs site and vacancies on the Fe sites. The second phase has a concentration of $Cs_{0.63(9)}Fe_{1.92(8)}Se_2$, which is consistent with Cs vacancies and filled Fe sites. The increased Fe concentration in the second phase is consistent with the increase in the $c$-axis lattice parameter of this phase. R-factors of the two solutions were quite high compared with typical published values at $R_w = 13.07\%$ ($R = 9.64\%$) and $R_w = 24.47\%$ ($R = 16.78\%$) respectively, though their large values are attributable to the difficulty in determining accurate intensities from partially-overlapping reflections. 

% Additional paragraph added for Ref 2, comment 4. 20 March 2015
An alternative method of selecting the peaks is to compare the difference in d-spacing between the two components of each peak to the resolution function of SXD, specified in Fig. 6 of Keen \textit{et al.}\cite{Keen2006}. Omitting reflections where the difference in d-spacing between the two sets of lattice parameters are less than the resolution at the particular $2\theta$ value resulted in a similar final set of reflections to the omission of reflections with $|Q|<4$\AA$^{-1}$ and the refinement results were found to be consistent.%, as shown in Table \ref{tab:RTtwo}(c).

\subsection{Superstructures}

Weaker peaks were observed away from the principal diffraction spots which can be described as commensurate superlattice peaks. The first set of peaks can be indexed on a commensurate grid defined by the lattice parameters from phase 1 and has propagation vectors:

\begin{align*}
	\vec{a}_{\sqrt 5} &= 2\vec{a}_1 - \vec{b}_1 \\
	\vec{b}_{\sqrt 5} &= \vec{a}_1 + 2\vec{b}_1\\
	\vec{c}_{\sqrt 5} &= \vec{c}_1
\end{align*}

This phase is named the \rtF{} phase, on account of the in-plane lattice vectors having a length of $a_1\sqrt{5}$. These propagation vectors determine one of two symmetrically equivalent domains, where the second domain can be generated by a reflection in either the $\{1,0,0\}$ or $\{1,1,0\}$ planes. The second set of superlattice peaks can only be indexed by lattice parameters of the second phase and has propagation vectors:

\begin{align*}
	\vec{a}_{\sqrt 2} &= \vec{a}_2 - \vec{b}_2 \\
	\vec{b}_{\sqrt 2} &= \vec{a}_2 + \vec{b}_2 \\
	\vec{c}_{\sqrt 2} &= \vec{c}_2
\end{align*}

This phase is named the \rtT{} phase, on account of the in-plane lattice vectors having a length of $a_2\sqrt{2}$. The superlattice peaks observed in this lattice are not peaked out of the plane but were instead found to exhibit rods along $l$, as shown in Fig. \ref{fig:SXDhhl}. This scattering was very weak and it was not possible to determine any intensity variation along the rod. Such 1D scattering occurs from a lack of correlation in the direction of the rod, implying that successive planes are not ordered with respect to one another.

% Section 4
\section{Synchrotron X-Ray Diffraction}
\begin{figure}[ht]
	\centering
	\includegraphics[width=0.5\textwidth]{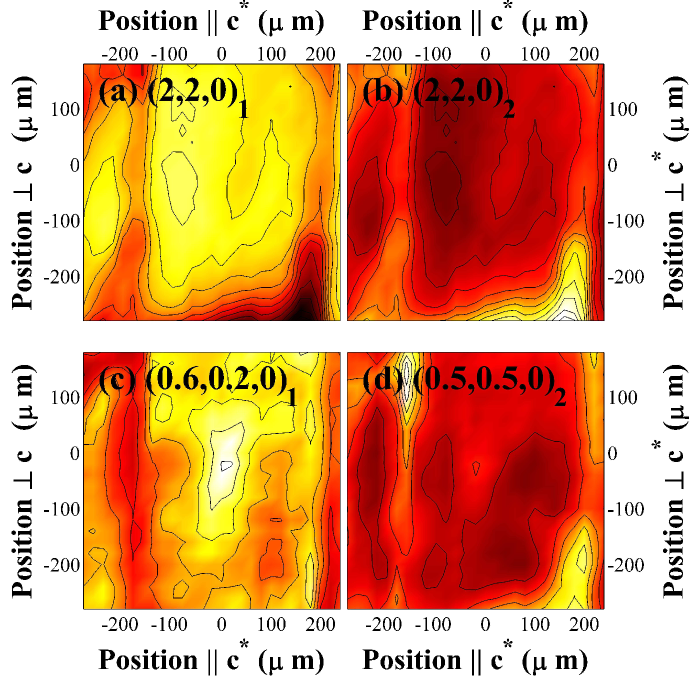}
	\caption{(Colour online) Scans across the surface of the sample at the (a) $(2,2,0)_1$, (b) $(2,2,0)_2$, (c) $(2.6,2.2,0)_1$ and (d) $(2.5,2.5,-1)_2$ reflections, where the reflections are defined using lattice parameters from phase 1 or 2. Each element is normalised against the sample surface by dividing by the sum of the two $(2,2,0)$ reflections.} \label{fig:FigI16Domains}
\end{figure}

Scans over a freshly cleaved surface were performed on I16 to determine the pattern of structural domains. The sample was mounted in reflection geometry in a cryostat and measurements were performed at the base temperature of 8K. The beam size was closed down to 20x10$\mu m$ and an energy of 7.05keV was chosen to make the x-ray penetration depth comparable to the beam size, with an attenuation length of $\approx 20 \mu m$. Scans across the sample were performed by aligning on a specific peak and then translating the sample through the beam. The Pilatus area detector was used to integrate over the peak in $2\theta$ and $\chi$, and a rocking scan in $\theta$ was performed at each point. Scans were taken at the two components of the principal diffraction spot and also at representative superlattice peaks for each phase. The results, shown in Fig. \ref{fig:FigI16Domains}, indicate that the regions of greatest intensity in phase 1 coincide with the regions of lowest intensity in phase 2. The same conclusion is drawn from the superlattice peaks, with the \rtF{} peak more intense in the regions where the \rtT{} peak is less intense, and vice versa. The size of these regions is roughly 300$\mu m$ x 300$\mu m$ for the dominant first phase and 50$\mu m$ x 300$\mu m$ for the minority phase.

% Figure 4

% Section 5
\section{Models}
Two separate phases can be distinguished in the experimental data, with different commensurate superlattices. The first phase has been well studied previously in the literature and can be described by a \rtF{} iron vacancy superstructure, illustrated in Fig \ref{fig:SXDhk0}(b). Simulations of this model are consistent with our diffraction data, shown in Fig. \ref{fig:SXDhk0}(a). The neutron refinements, which separate the two phases, indicate that the Cs occupancy is filled in this phase and indeed, simulating the diffracted intensities in each case indicates the relative intensities of the superlattice peaks do not change significantly when changing the Cs concentration.

\begin{figure}[ht]
	\centering
	\includegraphics[width=0.5\textwidth]{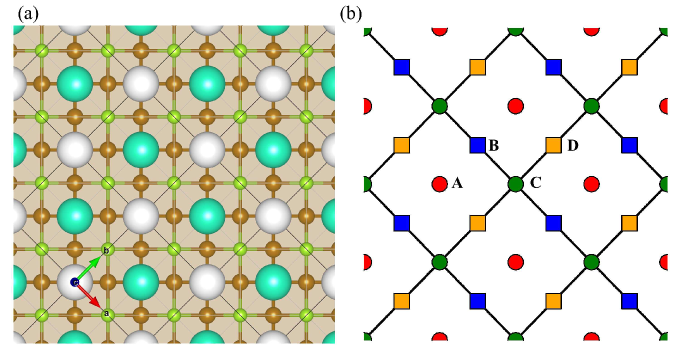}
	\caption{(Colour online) Cs layer ordering in the \rtT{} phase. Image (a) illustrates a single Cs layer, where Cs ions are light blue and vacancies are white. They are shown above octahedra formed by Fe and Se ions, coloured brown and green respectively. The diagram in (b) is a representation of the different possible layouts of Cs ions in different layers, green and red circles indicate the two possible layouts in the first layer and blue and orange squares indicate positions in the second layer. The small energy difference between these different choices leads to a lack of correlations between each layer, producing Bragg rods in the diffraction pattern.} \label{fig:FigStructures}
\end{figure}

Phase 2 has been observed previously in the literature as an unknown \rtT{} phase\cite{Ricci2011a,Bosak2012,Cai2012}. Our refinements indicate that only the Cs site contains vacancies in this phase. The commensurate supercell comprises two average unit cells, or two occupiable Cs sites per layer. The Cs ions sit at high symmetry positions in the parent space group, preventing their movement. The only solution that gives peak positions consistent with the diffraction pattern is to remove every other Cs ion from the layer as illustrated in Fig. \ref{fig:FigStructures}. This produces stripes of Cs ions and a concentration of $50\%$, which is consistent with the refinements. On consecutive Cs layers, there are two possible positions to place the vacancy arising from two layout options per layer, see Fig. \ref{fig:FigStructures}(b). It is not possible to determine, by symmetry, which layout should be preferred. Simulations of this model, expanded in the $c$ direction over a large number of cells with random orientations of Cs layers produce a diffraction pattern with 1D diffuse stripes along $l$, consistent with the neutron diffraction data, see Fig. \ref{fig:SXDhhl}.

% Figure 5

% Section 6
\section{\textit{Ab initio} calculations}
\textit{Ab initio} calculations were performed of the \rtT{} structure using the CASTEP code. To show that the proposed Cs vacancy structure is the origin of the 2D order it is necessary to show two things. First it must be shown that the system will order at the very least in-plane and second that there is no preferred stacking sequence. Two calculations were performed with different in-plane models, the first with the proposed ordered vacancy structure shown in Fig. \ref{fig:FigStructures}(a) and the second with a single Cs displaced to an unfavourable site. To investigate if the system will order 3 dimensionally it is then necessary to consider stacking sequences. The vacancy placement in the first two Cs layers is irrelevant as all four options, AB, AD, CD or CB are invariant under either translation or inversion symmetry. The third and fourth layers however can either sit directly on top of their respective original layers or in the previously vacant sites, having a stacking sequence such as ABAB or ABCD. These different options are illustrated in Fig. \ref{fig:FigStructures}(b). 

Within an expanded supercell of the Cs superstructure, with $\vec{a'} = 4\vec{a}_2$, $\vec{b'} = 4\vec{b}_2$ and $\vec{c'} = \vec{c}_2$, a single Cs ion was moved to a vacancy site. This movement led to an increase in total energy of $0.851$ eV. This is a large increase, though most of it is attributable to simple Coulombic repulsion. In the proposed structure each Cs has four nearest neighbours and a Cs-Cs bond length of $5.36$ \AA. The displaced Cs ion has 3 nearest neighbours and a bond length of $3.8$ \AA{} which from simple electrostatics gives an increase of $0.62$ eV. The remaining energy changes are likely due to next nearest neighbours and subtle changes in the magnetism of the system.

The two different stacking sequences proved to be extremely close in energy. The total energy difference between the two models was $1.5$ meV. This is extremely small and only a little larger than the convergence of the calculation. This difference in energy equates to an ordering temperature of $\approx20K$, implying that above this temperature there will be no correlations between subsequent layers, in agreement with the simulations of the Bragg rods above. At such a low ordering temperature, the large Cs ions are unlikely to be mobile and therefore the inter-layer disorder is likely to freeze in, explaining why no ordered version of this phase has been observed at low temperature.

% Section 7
\section{Discussion}
% Added reference Wang2011b.
There have been a large number of studies on the phase-separation in this class of systems, with microscopic phase separation being identified early on as a potential solution to the observed coexistence of superconductivity and magnetic ordering in the system\cite{Johnston2010}. Multiple phases have been observed in all the variations of $A_xFe_{2-y}Se_2$ with $(A=K,Rb,Cs)$ using TEM\cite{Wang2011b, Speller2012}, STM\cite{Li2011,Cai2012}, NMR\cite{Texier2012} and diffraction techniques\cite{Ricci2011,Ricci2011a,Wang2011,Pomjakushin2011,Bosak2012,YuPomjakushin2012}. In each case, the system is found to form two, spatially separated phases, where the first, dominant, phase is identified as the \rtF{} antiferromagnetic phase. The full structure of the second phase is, until now, unidentified. The size of these regions is smaller than that reported here, though STM and TEM techniques measure at a smaller length scale, implying that there is a patterning within the structural domains that we have measured. A refinement of the second phase was performed from powder neutron diffraction measurements by Pomjakushin \textit{et al.} giving a composition in $I4/mmm$ of $Rb_{0.61(5)}Fe_{2.2(1)}Se_2$, which is consistent with agreement with our refinements and model\cite{YuPomjakushin2012}.

Synchrotron x-ray analysis by Ricci \textit{et al.} established the presence of a \rtT{} periodic superstructure\cite{Ricci2011a} and Bosak \textit{et al.} observed the 1D rods and a monoclinic distortion\cite{Bosak2012} in this phase. These rods are consistent with our measurements, however we did not observe a similar monoclinic distortion on SXD due to the coarse wavevector resolution of the detectors. During our own synchrotron measurement we did observe a splitting of the \rtT{} peaks consistent with the monoclinic distortion, however for our surface scans we integrated over the different components. It was also noted in the report by Bosak \textit{et al.} that Phase 2 is I centred, precluding disorder in the FeSe layer stacking and the postulation was made for uncorrelated Cs layers\cite{Bosak2012}.

In the case of $Rb_xFe_{2-y}Se_2$, polarised neutron scattering has been used to determine that the \rtT{} diffuse rods have no magnetic scattering\cite{Wang2011}. NMR data by Texier \textit{et al.} on $Rb_xFe_{2-y}Se_2$ indicates the presence of a majority magnetic phase in the presence of a minority, Fe vacancy free superconducting phase\cite{Texier2012}. By probing the superconducting band gap in $K_xFe_{2-y}Se_2$ with a STM, Li \textit{et al.} determined that Fe vacancies are destructive to superconductivity in the system\cite{Li2011}, in agreement with the filled Fe layer in our model. Further analysis of STM images by Cai \textit{et al.} has suggested that the \rtT{} phase contains Se charge ordering\cite{Cai2012}. On its own however, such charge ordering could not account for the Bragg rods observed here as such a signal would be vanishingly weak using neutron diffraction. It is possible, however, that charge ordering is driven by in-plane ordering of the alkaline ions. The previous literature is therefore in good agreement with our Cs vacancy model for the secondary, superconducting phase.

A recent soft chemistry study by Sun \textit{et al.} has concluded that the highest superconducting temperatures in FeSe systems are achieved when the iron oxidation state is reduced below $+2$ and when the iron vacancy concentration is low\cite{Sun2015}. This is consistent with the lack of vacancies and reduced oxidation state implicit in our model with nominal concentration $Cs_{0.5}Fe_2Se_2$.

% Section 8
\section{Conclusion}
Single crystal neutron Laue diffraction and synchrotron x-ray diffraction have been combined to determine the separation of phases in \csfese{}. Refinements of the two overlapping phases indicate that one phase comprises Fe vacancies and the other Cs vacancies. The two phases have different superlattices, with the Fe vacancy phase having a \rtF{} superlattice and the Cs vacancy phase exhibiting a \rtT{} superlattice with Bragg rods along $l$. The two phases were determined to be spatially separated by x-ray diffraction measurements across the surface of the sample. The first phase can be attributed to the Fe vacancy phase well studied in the literature, however the Cs vacancy phase can now be determined as a $50\%$ Cs phase with Cs ions sitting at every other allowed site. First principles calculations were able to determine that while the 2D layers are stable, there are no correlations between the inter-layer ordering of the Cs ions, giving rise to 1D Bragg rods in reciprocal space. The observation of this second, spatially separated phase in the presence of a primary phase known to be highly magnetically ordered suggests that this new phase is contributing to the observed superconductivity of the system.

\begin{acknowledgments}
We are grateful for the financial support and hospitality of ISIS and Diamond Light Source Ltd. We thank the EPSRC for support through grants EP/J011150/1 and EP/J012912/1 and for use of the UK national supercomputing facility ARCHER under grant EP/F036809/1. Other calculations used STFC's e-Science facility. Images of atomic structures were created using the computer program VESTA, by Koichi Momma and Fujio Izumi\cite{Momma2011}. 
\end{acknowledgments}

% Create the reference section using BibTeX:
\bibliography{FeSuperconductor}

\end{document}